\documentclass[twocolumn,prl,a4paper,showpacs,floatfix,preprintnumbers,amsmath,amssymb]{revtex4}

\usepackage{exscale}
\usepackage{graphicx}
\usepackage{amsmath}
\usepackage{latexsym}
\usepackage{amsfonts}
\usepackage{amssymb}
\usepackage{amscd}
\usepackage{bm}           
\usepackage{bbm}

\newcommand{\ket}[1]{\ensuremath{|#1\rangle}}

\newcommand{\ketbra}[1]{\ensuremath{| #1 \rangle \langle #1 |}}

\newcommand{\Eins}{\ensuremath{\mathbbm 1}}

\newcommand{\BE}{\begin{equation}}
\newcommand{\EE}{\end{equation}}
\newcommand{\be}{\begin{equation}}
\newcommand{\ee}{\end{equation}}
\newcommand{\bea}{\begin{eqnarray}}
\newcommand{\eea}{\end{eqnarray}}
\newcommand{\bean}{\begin{eqnarray*}}
\newcommand{\eean}{\end{eqnarray*}}
\newcommand{\kommentar}[1]{}

\newcommand{\mean}[1]{\ensuremath{\langle #1 \rangle}}

\newcommand{\proj}[1]{\ketbra{#1}}
\newcommand{\tr}{{\rm Tr}}

\newcommand{\bc}{\begin{center}}
\newcommand{\ec}{\end{center}}

\newcommand{\ud}{\mathrm{d}}

\begin{document}

\title{Entanglement and sensitivity in precision measurements\\
with states of a fluctuating number of particles}

\author{P. Hyllus$^1$, L. Pezz{\'e}$^2$ and A. Smerzi$^1$}
\affiliation{
$^1$INO-CNR BEC Center and Dipartimento di Fisica, Universit{\`a} di Trento, I-38123 Povo, Italy \\
$^2$
Laboratoire Charles Fabry de l'Institut d'Optique, CNRS and Univ. Paris-Sud, Campus Polytechnique, RD 128, 
F-91127 Palaiseau cedex, France
}

\begin{abstract}

The concepts of separability, entanglement, spin-squeezing and Heisenberg limit
are central in the theory of quantum enhanced metrology.
In the current literature, these are well established only in the case of linear 
interferometers operating with input quantum states of a known fixed number of particles.
This manuscript generalizes these concepts and extends the quantum phase estimation theory by taking into account 
classical and quantum fluctuations of the particle number.
Our analysis concerns most of the current experiments on precision measurements
where the number of particles is known only in average. 
 
\end{abstract}

\date{\today}


\maketitle

{\em Introduction}.
After the first pioneer works of the 80's and early 90's \cite{Caves_1981,80works,Wineland_1994,Kitagawa_1993},
the field of quantum-enhanced metrology is witnessing a revival of interest
due to possible implications on fundamental questions of quantum information 
and technological applications. 
Recent theoretical analyses are mainly focusing on the 
interplay between spin-squeezing, entanglement and phase sensitivity \cite{Pezze_2009,Sorensen_2001},
the set of fundamental sensitivity bounds \cite{Giovannetti_2006,Hofmann_2009,RivasArXiv10}
and the quest for optimal phase estimation protocols \cite{Pezze_2008, Huang_2008}.
This activity has came along with breakthrough
experimental investigations with photons \cite{Photons}, ions \cite{Ions}, cold atoms
\cite{ColdAtoms} and Bose-Einstein condensates (BECs) \cite{BEC, Bohi_2009}.
The interplay between theory and experiments is playing a crucial role for the development
of the field. 
It should be noticed, however, that while the theoretical investigations
have been developed in the context of systems having a fixed, known, number of particles,
most experiments have been performed with a fluctuating number of particles.
The consequences of classical and quantum fluctuations have been generally disregarded. 

The goal of this manuscript is to extend along this direction the quantum metrology theory, 
and, in particular, to discuss the relation between
{\it separability, entanglement}, 
{\it spin-squeezing} and {\it sub shot-noise sensitivity}, and 
to settle the concept of the {\it Heisenberg limit}.
In this framework, we recognize as entangled, for instance, the state suggested by Caves 
in '81 in the context of Mach-Zehnder interferometry
\cite{Caves_1981}, which provided the first proposal for sub shot-noise
phase estimation.

{\em Separability and Entanglement}. 
The fluctuation of the total number of particles can have a classical
and/or quantum nature. It is generally believed that quantum coherences
between states of different numbers of particles do not play any observable
role because of the existence of superselection rules (SSR) for the total number
of particles \cite{Sanders_2003,WickPR52,Bartlett_2007}. Therefore, the most general states can be written
as 
\be\label{eq:incoherent}
\hat\rho_\mathrm{inc}=\sum_{N=0}^{+\infty} Q_N \hat \rho^{(N)},
\ee 
where $\hat\rho^{(N)}$ is a state of $N$ 
particles. 
We extend the usual definition of separability of states
of $N$ particles \cite{WernerPRA89}
by considering such incoherent mixtures
as separable if they can be written as
\be\label{sep}
\hat\rho_{\rm sep}=\sum_{N=0}^{+\infty} Q_N \hat\rho^{(N)}_{\rm sep},
\ee
where 
$\hat{\rho}^{(N)}_{\rm sep}=\sum_k P_{k,N} \proj{\phi_{k,N}^{(1)}} \otimes \cdots \otimes \proj{\phi_{k,N}^{(N)}}$,
$\{Q_N\}$ and $\{ P_{k,N} \}$ are probability distributions and
$\ket{\phi_{k,N}^{(j)}}$ is a two-mode pure state of a single particle \cite{nota_sep}. 
The existence of a SSR is the consequence of the lack of a suitable
phase reference frame (RF) \cite{Bartlett_2007}. 
However, the possibility that a suitable RF can be established 
in principle cannot be excluded \cite{Bartlett_2007}.
If SSRs are lifted, then states $\hat \rho_\mathrm{coh}$ containing coherent superpositions 
of different numbers of particles would become physically relevant.
These will be called separable if they are separable in every fixed-$N$ subspace, 
{\em i.e.}, if the state $\hat \rho_{\rm red} \equiv \sum_N \Eins_N\hat\rho_\mathrm{coh}\Eins_N$
without such coherences is of the form of Eq.~(\ref{sep}) \cite{nota_sep}.
States which are not separable according to this definition are entangled \cite{nota_disting}.
In this manuscript, it will be shown that entanglement is a 
necessary resource for sub shot-noise sensitivity in linear interferometers.

{\em Phase sensitivity with a linear interferometer}.
We consider a linear two-mode interferometer
where the density matrix $\hat \rho$ evolves as
$\hat \rho(\theta) = e^{-i \theta \hat J_{\vec n}} \hat \rho e^{+i \theta \hat J_{\vec n}}$, 
where $\theta$ is a real number, ${\vec n}$ is an 
arbitrary direction in the three dimensional space, 
and $\hat J_{\vec n} = \vec{\hat J} \cdot \vec n$ is a collective spin operator. 
For a non-fixed number of particles  $\vec{\hat J}$ 
is defined as 
$\vec{\hat J}=\oplus_{N=1}^{+\infty} \vec{\hat J}^{(N)}$,
where $\vec{\hat J}^{(N)} =\frac{1}{2}\sum_{l=1}^N \vec{\hat\sigma}^{(l)}$ and 
$\vec{\hat\sigma}^{(i)}$ is the vector of Pauli matrices acting on the $l$-th particle
\cite{Schwinger}.
Note that linear interferometers preserve the number
of particles, $[\hat{J}_{\vec n},\hat N]=0$.
The goal is to estimate the parameter $\theta$ 
with the maximum possible sensitivity $\Delta \theta$.
The quantum Cramer-Rao theorem ensures that
for arbitrary, unbiased, phase estimation protocols, the phase sensitivity 
is bounded by \cite{nota_Fisher, Braunstein_1994}
\be \label{QCR}
\Delta \theta_{\rm QCR} = \frac{1}{\sqrt{m F_Q[\hat{J}_{\vec n},\hat{\rho}]}}.
\ee
Here $m$ is the number of independent measurements done with
identical input states $\hat\rho$ and $F_Q[\hat{J}_{\vec n},\hat{\rho}]$
is the quantum Fisher information (QFI) \cite{nota_Fisher}.
Equation (\ref{QCR}) is a bound on $\Delta\theta$ given the state
$\hat\rho$ and the number of measurements $m$ and can be saturated 
in the central limit.

In the following, we set the fundamental sensitivity
bounds of the phase estimation problem 
by taking, as a constraint,    
the average total number of particles, 
$\bar N_{\rm tot}=m \langle \hat{N} \rangle$, without fixing $m$ or 
$\mean{\hat N}$ separately.
This correctly accounts for the finite available resources \cite{Braunstein_1992,Pezze_2008}.
Averages are computed with respect to
the input state of the interferometer.

{\em Shot-noise limit.}
We define the shot-noise limit as the maximal phase sensitity attainable with separable states.
For separable incoherent mixtures, Eq.~(\ref{sep}),
$F_Q[\hat\rho_{\rm sep},\hat J_{\vec n}]=\sum_N Q_N F_Q[\hat\rho_{\rm sep}^{(N)},\hat J_{\vec n}^{(N)}]
\le \sum_N Q_N N=\langle \hat N\rangle$. The first equality is due to the block-diagonal
form of $\hat J_{\vec n}$ and $\hat \rho_{\rm sep}$ \cite{nota_FQ} and the second inequality was proved
in Ref.~\cite{Pezze_2009}.
By using this result and Eq.~(\ref{QCR}),
the shot-noise limit is thus given by  
\begin{equation} \label{SN}
\Delta \theta_{\mathrm{SN}} = \frac{1}{\sqrt{\bar N_{\rm tot}}},
\end{equation}
which agrees with the common definition of the shot-noise or
standard quantum limit.
As shown in Appendix A, Eq.~(\ref{SN}) holds also 
when considering separable states with coherences. 
This brings us to the following results.

{\em Entanglement and sub shot-noise sensitivity}. 
An arbitrary state with non-fixed number of particles is entangled 
if it fulfills the inequality 
\be \label{Fisher}
	\chi^2 \equiv \frac{\langle \hat N \rangle}{F_Q[\hat{J}_{\vec n},\hat{\rho}]} <1,
\ee
for some direction $\vec n$. 
States satisfying Eq.~(\ref{Fisher}) are useful 
in a linear interferometer implemented by the transformation 
$\hat{J}_{\vec n}$, since, according to Eq.~(\ref{QCR}), 
they provide a sub shot-noise (SSN) phase sensitvity.
We further introduce the spin-squeezing condition
\be \label{SpinSqueezing}
\xi^2 \equiv \frac{ \langle \hat N \rangle (\Delta \hat{J}_{\vec n_3})^2 }
{ \langle \hat{J}_{\vec n_1} \rangle^2 + \langle \hat{J}_{\vec n_2} \rangle^2} < 1,
\ee
where the vectors $\vec n_1$, $\vec n_2$ and $\vec n_3$ define a right-handed
coordinate system \cite{CRs}.
In Appendix B, we prove that $\chi^2 \leq \xi^2$ holds.
Hence, spin-squeezed states ($\xi < 1$) are entangled and useful for SSN interferometry. 
Equations (\ref{Fisher}) and (\ref{SpinSqueezing}) generalize 
the conditions for entanglement and SSN discussed in 
\cite{Wineland_1994,Sorensen_2001,Pezze_2009} to states of non-fixed 
number of particles and only require the replacement of $N$ with $\langle \hat N \rangle$.
This result has been previously used without justification
in several experimental works.

{\em The Heisenberg limit.}
While it is well known that the shot-noise limit Eq.~(\ref{SN})
can be overcome, finding the Heisenberg limit (HL), {\em i.e.} the ultimate scaling of the 
phase sensitivity imposed by Quantum Mechanics for states with non-fixed $N$, has generated a vivid debate \cite{Shapiro_1989,Braunstein_1992,Ou_1996,Hofmann_2009,Anisimov_2009}.
For instance, it has been suggested that $\Delta \theta \sim 1/\langle \hat{N} \rangle$
is the fundamental sensitivity bound \cite{Ou_1996}.
Recently it was argued that 
the HL is given by $\Delta \theta = 1/\sqrt{\langle \hat{N}^2 \rangle}$ \cite{Hofmann_2009}.
Not only this limit overcomes the bound $\Delta \theta = 1/\langle \hat{N} \rangle$ 
but, since the quantity $\langle \hat{N}^2 \rangle$ can grow arbitrarily 
fast with $\langle \hat{N} \rangle$,
the phase sensitivity can be arbitrarily high when considering wildly fluctuating 
number distributions \cite{nota_SSW}. 

The definition of the HL depends on the specific
constraints imposed on the resources.
In the presence of the SSR, and by taking $\langle \hat{N} \rangle$, 
$\langle \hat{N}^2 \rangle$ and $m$ as constraints,
the phase sensitivity is bounded by
\be\label{HLinc}
\Delta\theta \geq \max\Big[\frac{1}{\sqrt{m\mean{\hat N^2}}},\frac{1}{m \mean{\hat N}}\Big].
\ee
The first bound is derived from Eq.~(\ref{QCR}) by noticing that, 
for incoherent mixtures~(\ref{eq:incoherent}),
$F_Q[\hat \rho_{\rm inc},\hat J_{\vec n}]=\sum_N Q_N F_Q[\hat\rho^{(N)},\hat J_{\vec n}^{(N)}]
\le\sum_N Q_N N^2=\mean{\hat N^2}$.
It agrees with the bound of \cite{Hofmann_2009}, except for the 
factor $m$, accounting for repeated independent measurements.
Note that in the central limit, the bound scales as $1/\sqrt{m}$.
The second bound is proven in Appendix C. 
When the number of particles is fixed and equal to $N$, Eq.~(\ref{HLinc})
recovers the definition of the HL $\Delta\theta = 1/ \sqrt{m}N$ 
discussed in \cite{Giovannetti_2006}
which takes, as constraints, $N$ and $m$, separately. 
Note, however, that from this
limit one does not obtain Eq.~(\ref{HLinc}) by naively replacing $N\to \mean{\hat N}$.

Here, we define the HL as the maximum phase sensitivity 
attainable with fixed $\bar N_{\rm tot}$. 
If coherences can neither be generated nor measured,
as in current experimental interferometric protocols because of the SSR, the HL is  
\be\label{HL_tot}
\Delta\theta_{\rm HL} = \frac{1}{\bar N_{\rm tot}}.
\ee 
If the state contains coherences but only POVMs without number-coherences
are applied, the HL is still given by Eq.~(\ref{HL_tot}), cf. Appendix C. 
In particular, this is true for the Mach-Zehnder interferometer, 
even if only 
the relative 
number of particles is measured or the parity at one exit port.
The latter measurement was considered in Ref.~\cite{Anisimov_2009}.
By adapting the proof in Appendix C it can also be shown that the HL Eq.~(\ref{HL_tot}) holds
for incoherent mixtures even if POVMs with 
coherences are available.

If the state contains coherences and POVMs with coherences are used, then
Eq.~(\ref{HL_tot}) is valid in the central limit only \cite{nota_HL}.
Outside the central limit the bound
\be\label{HLcoh}
\Delta\theta \geq \frac{1}{\sqrt{m\langle \hat N^2\rangle}}
\ee
holds, as shown in Appendix C.
However, there are arguments suggesting that the HL may be given by Eq.~(\ref{HL_tot}) also in this case
\cite{LevitinPRL09,Braunstein_1992}.

{\em Examples.} We consider a Mach-Zehnder interferometer, 
corresponding to a collective rotation around the $y$ axis, {\em i.e.}, $\vec n=\hat y$,
with a coherent state $|\alpha\rangle_a$ in the input mode $a$
and a squeezed state $|\zeta\rangle_b$ in the input $b$ 
as suggested by Caves \cite{Caves_1981,Pezze_2008,Nota_Caves}.
Here $\alpha = |\alpha|e^{i \phi_\alpha}$ and 
$\zeta=r e^{i \theta_\zeta}$. 
When fixing the relative phase $2\phi_\alpha=\theta$, Eq.~(\ref{Fisher}) gives
\be \label{chiSV}
\chi^2 = \frac{|\alpha|^2 + \sinh^2 r}
{|\alpha|^2 e^{2r} + \sinh^2 r}.
\ee
For $r=0$ we have a coherent state entering one of the two ports and vacuum
the other port. 
We obtain $\chi^2=1$ and hence shot noise, as expected.
In contrast, we obtain $\chi^2<1$ when $r>0$ and $|\alpha|^2 \neq 0$.
The state is entangled and, 
for relatively small values of $r$,
it is also spin squeezed, according to the definition Eq.~(\ref{SpinSqueezing}).
A direct calculation for ${\vec n}_3=\hat x$
gives 
\be \label{xiSV}
\xi^2 = \frac{(|\alpha|^2 + \sinh^2 r)(|\alpha|^2 e^{-2r}+ \sinh^2 r)}{(|\alpha|^2 - \sinh^2 r)^2}.
\ee
In the limit $|\alpha|^2 \gg \sinh^2 r$ we have that $\xi^2 = e^{-2 r} < 1$ 
for $r>0$ (and $|\alpha| \neq 0$). 
Spin squeezing is lost for $|\alpha|^2 \sim \sinh^2 r$, even though
the state remains entangled.
Eqs~(\ref{chiSV},\ref{xiSV}) hold even for the state 
$\rho_{\rm red}=\sum_N \Eins_N |\alpha,\zeta\rangle\langle\alpha,\zeta|\Eins_N$ obtained in
the presence of the SSR \cite{Bartlett_2007,nota_ex}. 
We finally notice that the state $|\alpha\rangle \otimes |\zeta \rangle$ 
reaches the HL $\Delta \theta \sim 1/\bar N_{\rm tot}$ 
when  $|\alpha|^2 \approx \sinh^2 r$ \cite{Pezze_2008}.

{\em Conclusions}.
In this manuscript we have extended the quantum phase estimation theory to the case of 
states with fluctuating number of particles. In particular, we have
discussed entanglement and spin-squeezing and provided the class of entangled states
useful for sub shot-noise sensitivity. The Heisenberg limit
has been defined by taking into account the finite available 
resources. Our results are relevant for most current experiments on the creation,
detection and application of entangled states with a fluctuating number
of particles for quantum metrology. 

{\em Appendix A. Shot-noise limit}. 
Here we demonstrate that $F_Q(\hat\rho_{\rm cs}) \leq \langle \hat N \rangle$
for states of the form $\hat\rho_{\rm cs}=\sum_k p_k \proj{\psi_{\rm cs}^{(k)}}$,
which contain coherences between different numbers of particles 
and are separable in every fixed-$N$ subspace.
We obtain
$F_Q(\hat\rho_{\rm cs})\le $  
 max $4\sum_k p_k (\Delta \hat J_{\vec n})^2_{\ket{\psi_{\rm cs}^{(k)}}}
\le 4\sum_k p_k \sum_N Q_{N}^{(k)}$ max $\langle [\hat J_{\vec n}^{(N)}]^2\rangle_{\ket{\psi_{\rm cs}^{(N,k)}}}
\le \sum_k p_k \sum_N Q_N^{(k)} N = \langle \hat N \rangle$,
by using the convexity of the quantum Fisher information \cite{convexity}.
We expanded
$\ket{\psi_{\rm cs}^{(k)}}=\sum_N \sqrt{Q_N^{(k)}}\ket{\psi_{\rm cs}^{(N,k)}}$ 
and used that for pure separable states 
$\langle [\hat J_{\rm vec}^{(N)}]^2\rangle_{\ket{\psi_{\rm sep}^{(N)}}}\le
\frac{N}{4}$ \cite{Pezze_2009}. 
By inserting $F_Q(\hat\rho_{\rm cs}) \leq \langle \hat N \rangle$ into 
Eq.~(\ref{QCR}), we recover Eq.~(\ref{SN}).

{\em Appendix B. Spin-squeezing inequality}. 
We consider, without loss of generality, 
a coordinate system such that $\mean{\hat{J}_{\vec{n}_2}}=0$.
From the inequalities 
$F_Q[\hat \rho,\hat J_{\vec n}] \geq F_{\hat E(\varepsilon)}[\hat \rho(\theta)]$ 
\cite{nota_Fisher} and
$F_{\hat E(\varepsilon)}[\hat \rho(\theta)] \geq \frac{1}{(\Delta\hat A)^2} \big( \frac{\ud \langle \hat A\rangle}{\ud \theta} \big)^2$ (obtained by extending the proof in \cite{UysPRA07}
to the case of non-fixed number of particles), it follows that 
$\chi^2\le \langle \hat N \rangle (\Delta \hat A)^2 / (\ud \langle \hat A\rangle / \ud \theta)^2$.
By choosing $\vec{n} = \vec{n}_2$, $\hat A = \hat J_{\vec n_3}$ and using 
the commutation relations of the $\hat J_{\vec n_i}$ operators 
\cite{CRs}, we obtain $\frac{\partial}{\partial \theta} \langle \hat J_{\vec n_3}\rangle =
i\mathrm{Tr}\big[ \hat{J}_{\vec{n}_3}[\hat{J}_{\vec{n}_2}, \hat{\rho}(\theta)]\big] = 
\langle \hat{J}_{\vec{n}_1} \rangle$.
Hence $\chi^2 \le \frac{ (\Delta \hat{J}_{\vec{n}_3} )^2 }{\langle \hat{J}_{\vec{n}_1} \rangle^2} = \xi^2$.

{\em Appendix C. Heisenberg limit}. 
In the presence of the SSR, only POVMs without coherences are available.
A POVM 
is of this form if we can write its elements as 
$\hat E(\lambda)=\sum_{(N,M)\in I(\lambda)}\hat E_{N,M}$, where 
$\sum_M \hat E_{N,M}=\Eins_N$ 
($M$ is a degree of freedom in each $N$-subspace), and
$I(\lambda)$ are all pairs $(N,M)$ leading to the same $\lambda$
($\lambda$ could be the number of particles at one port of 
a Mach-Zehnder interferometer, for instance).
For $m=1$, the conditional probabilities are
$P(\lambda)=\sum_{(N,M)\in I(\lambda)}P(N,M|\theta)=\sum_{(N,M)\in I(\lambda)} Q_N P(M|\theta,N)$,
where $P(M|\theta,N)=\tr\big[\hat E_{N,M} e^{i \theta \hat J_{\vec n}}\hat \rho^{(N)} e^{-i \theta \hat J_{\vec n}}\big]$, and $Q_N \hat\rho^{(N)}=\Eins_N\hat\rho\Eins_N$. The variance of 
an estimator $\theta_{\rm est}(\lambda)$ (assumed to be unbiased, {\em i.e.}, $\bar \theta_{\rm est}=\theta$)
is $(\Delta\theta_{\rm est})^2 
=\sum_\lambda P(\lambda|\theta)(\theta_{\rm est}(\lambda)-\theta)^2
=\sum_\lambda \sum_{(N,M)\in I(\lambda)}Q_N P(M|\theta,N)(\theta_{\rm est}(\lambda)-\theta)^2
=\sum_N Q_N \sum_M P(M|N,\theta)
(\theta_{\rm est}(\lambda[N,M])-\theta)^2=
\sum_N Q_N(\Delta\theta_{\rm est}^{(N)})^2 
\ge \sum_N Q_N/N^2$. We used that $\sum_\lambda\sum_{(N,M)\in I(\lambda)}=\sum_N\sum_M$ holds
since $\sum_\lambda \hat E(\lambda)=\Eins$. In the last inequality we used that
$(\Delta\theta_{\rm est}^{(\vec N)})^2 \geq 1/N^2$ holds for unbiased estimators \cite{Giovannetti_2006}.
For $m\ge 1$, $P(\vec \lambda|\theta)=\prod_{i=1}^m P(\lambda_i|\theta)$
and it can be shown that
$
(\Delta\theta_{\rm est})^2\ge 
\sum_{\vec N}Q_{\vec N}/\sum_{i=1}^m N_i^2\ge \sum_{\vec N} Q_{\vec N}/(\sum_{i=1}^m N_i)^2,
$
where the first inequality follows as above and the second inequality holds for positive
numbers.
For $m=1$, we then obtain $(\sum_N Q_N/N^2)(\sum_N Q_N)\ge (\sum_N Q_N/N)^2$
from the Cauchy-Schwartz inequality. Applying it again leads to
$(\sum_N Q_N/N)(\sum_N Q_N N)\ge (\sum_N Q_N)^2=1$. 
Therefore, we arrive at
$\Delta \theta_{\rm est} \ge 1/\langle \hat N\rangle$.
This can be done in analogy for $m\ge 1$.
This proves that $\Delta\theta_{\rm est} \geq 1/m\mean{\hat N}$.

The bound~(\ref{HLcoh}) can be proven as in Appendix A by using
general states and $F_Q[\ket{\psi^{(N)}},\hat J_{\vec n}]\le N^2$. 

{\em Acknowledgements}. We thank J.I. Cirac, K. M{\o}lmer, A.S. S{\o}rensen and T. Rudolph
for stimulating discussions.



\begin{thebibliography}{99}

\bibitem{Caves_1981} C.M. Caves, \emph{Phys. Rev. D} {\bf 23}, 1693 (1981).

\bibitem{80works} 
B. Yurke, S.L. McCall and J.R. Klauder, \emph{Phys. Rev. A} \textbf{33}, 4033 (1986);
M. Xiao, L.A. Wu and H.J. Kimble, \emph{Phys. Rev. Lett.} \textbf{59}, 278 (1987);
P. Grangier, R.E. Slusher, B. Yurke and A. LaPorta, \emph{Phys. Rev. Lett.} \textbf{59}, 2153 (1987);
M.J. Holland and K. Burnett, \emph{Phys. Rev. Lett.} \textbf{71}, 1355 (1993).

\bibitem{Wineland_1994} D.J. Wineland, 
J.J. Bollinger, W.M. Itano, and D.J. Heinzen, \emph{Phys. Rev. A} \textbf{50}, 67 (1994).

\bibitem{Kitagawa_1993} M. Kitagawa and M. Ueda, \emph{Phys. Rev. A} \textbf{47}, 5138 (1993).

\bibitem{Sorensen_2001} 
A. S\o rensen, {\em et al.}, \emph{Nature} \textbf{409}, 63 (2001).

\bibitem{Pezze_2009} L. Pezz\'e and A. Smerzi, \emph{Phys. Rev. Lett.} \textbf{102}, 100401 (2009). 

\bibitem{Giovannetti_2006} V. Giovannetti, S. Lloyd and L. Maccone, 
\emph{Phys. Rev. Lett.} \textbf{96}, 010401 (2006).

\bibitem{Hofmann_2009} H.F. Hofmann, \emph{Phys. Rev. A} \textbf{79}, 033822 (2009).

\bibitem{RivasArXiv10}
{\'A}. Rivas and A. Luis, Phys. Rev. Lett. {\bf 105}, 010403 (2010). 

\bibitem{Pezze_2008} L. Pezz\'e and A. Smerzi, \emph{Phys. Rev. Lett.} \textbf{100}, 073601 (2008).

\bibitem{Huang_2008} Y.P. Huang and M.G. Moore, \emph{Phys. Rev. Lett.} \textbf{100}, 250406 (2008);
U. Dorner, \emph{et al.}, \emph{Phys. Rev. Lett.} \textbf{102}, 040403 (2009);
D.W. Berry, \emph{et al.}, \emph{Phys. Rev. A} \textbf{80}, 052114 (2009);

\bibitem{Photons} 
P. Walther, {\em et al.}, \emph{Nature} \textbf{429}, 158 (2004); 
M.W. Mitchell, {\em et al.}, \emph{Nature} \textbf{429}, 161 (2004);
T. Nagata, {\em et al.}, \emph{Science} \textbf{316}, 726 (2007).

\bibitem{Ions} 
V. Meyer {\em et al.}, \emph{Phys. Rev. Lett.} \textbf{86}, 5870 (2001); 
D. Leibfried {\em et al.}, {\em Science} {\bf 304}, 1476 (2004).

\bibitem{ColdAtoms}
J. Appel, 
et al., \emph{PNAS} \textbf{106}, 10960 (2009);
M.H. Schleier-Smith, I.D. Leroux, V. Vuletic,
\emph{Phys. Rev. Lett.} \textbf{104}, 073604 (2010). 

\bibitem{BEC} 
G.B. Jo {\em et al.}, \emph{Phys. Rev. Lett.} \textbf{98}, 030407 (2007);
J. Est{\`e}ve, {\em et al.}, \emph{Nature} \textbf{455}, 1216 (2008).

\bibitem{Bohi_2009} 
P. B\"ohi, {\em et al.}, \emph{Nat. Phys} \textbf{5}, 592 (2009).

\bibitem{WickPR52}
G.C. Wick, \emph{et al.}, \emph{Phys. Rev.} {\bf 88}, 101 (1952).


\bibitem{Sanders_2003} 
B.C. Sanders, S.D. Bartlett, T. Rudolph and P.L. Knight, \emph{Phys. Rev. A} \textbf{68}, 042329 (2003); 
K. M{\o}lmer, \emph{Phys. Rev. A} \textbf{55}, 3195 (1997);


\bibitem{Bartlett_2007}
S.D. Bartlett {\em et al.}
{\em Rev. Mod. Phys.} {\bf 79}, 555 (2007).

\bibitem{WernerPRA89}
R.F. Werner, {\em Phys. Rev. A} {\bf 40}, 4277 (1989).

\bibitem{nota_sep}
Density matrices $\hat\rho^{(N)}$ of $N$ particles 
with
$2$ degrees of freedom act on a Hilbert space 
${\cal H}_N=\otimes_{j=1}^N {\cal H}^{(j)}_N$,
where ${\cal H}_N^{(j)}\simeq\mathbb{C}^2$. 
The operator $\Eins_N$ is the projector onto ${\cal H}_N$.

\bibitem{nota_disting}
The definition of separability can also be applied 
formally when treating indistinguishable particles in first
quantization, where the set of admissible quantum states is reduced to the 
symmetric/antisymmetric subspace for bosons/fermions.
Separable states of bosons without coherences are then 
given by Eq.~(\ref{sep}) with $\ket{\phi_{k,N}^{(i)}}=\ket{\phi_{k,N}}$ 
for all $i$. It has been argued that particle entanglement 
due to (anti-)symmetrization in first quantization
is unphysical since the individual particles cannot be addressed
\cite{ZanardiPRA02}.
However, in interferometers using only collective
operations, it can still be a useful resource for sub 
shot-noise sensitivity.

\bibitem{ZanardiPRA02}
P. Zanardi, {\em Phys. Rev. A} {\bf 65}, 042101 (2002);
A. P. Hines, R. H. McKenzie, and G. J. Milburn,
{\em Phys. Rev. A} {\bf 67}, 013609 (2003);
F. Benatti, R. Floreanini, and U. Marzolino, {\em Ann. Phys.} {\bf 325}, 
924 (2010).
See also 
L. Amico {\em et al.}, {\em Rev. Mod. Phys.} {\bf 80}, 517 (2008)
and References within.

\bibitem{Schwinger} 
For bosons, 
the operators ${\hat J_{x,y,z}}$ can be written as 
$\hat J_x=(\hat a^\dagger\hat b+\hat b^\dagger\hat a)/2$,
$\hat J_y=(\hat a^\dagger\hat b-\hat b^\dagger\hat a)/2i$,
and $\hat J_z=(\hat a^\dagger\hat a-\hat b^\dagger\hat b)/2$
using annihilation operators $\hat a$ and $\hat b$ 
for the two interferometric modes.

\bibitem{Braunstein_1994} 
S.L. Braunstein and C.M. Caves, \emph{Phys. Rev. Lett.} \textbf{72}, 3439 (1994).

\bibitem{nota_Fisher} To estimate the parameter $\theta$, a positive operator value measurement 
(POVM), $\{\hat E(\varepsilon)\}$, is performed on the output state and 
$\theta$ is inferred from the result of this measurement.
The maximum sensitivity of this estimation is 
bounded by the Cramer-Rao bound $\Delta \theta_{\rm CR} = 1/ \sqrt{m F_{\hat E(\varepsilon)}[\hat\rho(\theta)]}$, 
where $m$ is the number of measurements, $F_{\hat E(\varepsilon)}[\hat\rho(\theta)]
=\int d\varepsilon P(\varepsilon|\theta)(\partial_\theta \log P(\varepsilon|\theta)/\partial \theta)^2$ is the Fisher information
and $P(\varepsilon|\theta)=\tr[\hat E(\varepsilon) \hat\rho(\theta)]$.
The QFI is the maximum 
$F_Q[\hat\rho(\theta)] \equiv \max_{\{\hat E(\varepsilon)\}}F_{\hat E(\varepsilon)}[\hat\rho(\theta)]$
and it is saturated by a particular POVM \cite{Braunstein_1994}.



\bibitem{Braunstein_1992} 
S.L. Braunstein, A.S. Lane and C.M. Caves, \emph{Phys. Rev. Lett.} \textbf{69}, 2153 (1992);
A.S. Lane, S.L. Braunstein and C.M. Caves, \emph{Phys. Rev. A} \textbf{47}, 1667 (1993).


\bibitem{nota_FQ}
The QFI for a state of fixed number of particles, 
$\hat \rho^{(N)}$, is $F_\mathrm{Q} [\hat{\rho}^{(N)},\hat{J}_{\vec{n}}^{(N)}] = 4 (\Delta \hat{R}^{(N)})^2$, 
where the Hermitean operator $\hat{R}^{(N)}$ satisfies the equation
$\{ \hat{R}^{(N)}, \hat{\rho}^{(N)} \}= i [\hat{J}_{\vec{n}}^{(N)}, \hat{\rho}^{(N)}]$.
For pure states, $F_Q[\ket{\psi},\hat J_{\vec n}]=4(\Delta\hat J_{\vec n})^2_{\ket{\psi}}$.
If the input state is an incoherent mixture $\hat \rho=\sum_N Q_N \hat \rho^{(N)}$,
then $F_\mathrm{Q} [\hat{\rho}, \hat{J}_{\vec{n}}]= \sum_N Q_N F_\mathrm{Q} [\hat{\rho}^{(N)}, \hat{J}_{\vec{n}}^{(N)}]$ 
follows since $\hat\rho$ and $\hat J_{\vec n}$ commute with $\hat N$, which can also be imposed for $\hat R$. 
The POVM saturating $F=F_Q$ 
can be chosen as a von Neumann measurement on the eigenstates of $\hat R$ \cite{Braunstein_1994},
therefore a POVM without coherences 
saturates $F_Q$.
However, for non-restricted states and POVMs
$F_Q(\hat\rho)\ge F_Q(\sum_N \Eins_N\hat\rho\Eins_N)$ holds.
A simple example where the inequality is strict is
$(\ket{y+}+\ket{y-}\otimes\ket{z+})/\sqrt{2}$ for the generator $\hat J_y$,
where $\ket{(y,z)\pm}$ are the eigenstates of $\hat\sigma_{y,z}$.

\bibitem{CRs}
For a right-handed coordinate system ${\vec n}_i$, $i=1,2,3$, 
the operators $\hat J_{\vec n_i}$ satisfy the SU(2)
algebra of angular momentum: $[\hat J_{\vec n_i}, \hat J_{\vec n_j}] = i \epsilon_{i,j,k} \hat J_{\vec n_k}$, 
where $\epsilon_{i,j,k}$ is the Levi-Civita tensor.



\bibitem{Ou_1996} Z.Y. Ou, 
\emph{Phys. Rev. A} \textbf{55}, 2598 (1997).


\bibitem{Anisimov_2009} P.M. Anisimov, \emph{et al.}, \emph{Phys. Rev. Lett.} \textbf{104}, 103602 (2010).

\bibitem{Shapiro_1989} J.H. Shapiro, S.R. Shepard and N.C. Wong, 
\emph{Phys. Rev. Lett.} \textbf{62}, 2377 (1989).

\bibitem{nota_SSW}
This happens, for instance, for the state proposed by
Shapiro, Shepard and Wong (SSW) \cite{Shapiro_1989}.
However, the SSW result did not take into account the number of measurements.
Indeed, it was shown that the SSW proposal does not overcome the limit 
$\Delta \theta = 1/\bar{N}_{\rm tot}$ \cite{Braunstein_1992}.

\bibitem{nota_HL} 
Fisher's theorem ensures that 
the bound $1/(m\mean{\hat N^2})^{1/2}$ can be saturated in the central limit
for $m\ge m_{\rm cl}$, where $m_{\rm cl}$ depends on the POVM used. 
In particular, it can be saturated with POVMs without coherences.
In this case, and for $m\ge m_{\rm cl}^{\rm inc}$, $1/(m\mean{\hat N^2})^{1/2}\ge 1/m\mean{\hat N}$ holds due to Eq.~(\ref{HLinc}). 
However, for $m'<m_{\rm cl}^{\rm inc}$, we cannot rule out
that $\Delta\theta < 1/\bar N_{\rm tot}$ can be achieved if 
states and POVMs with coherences are available.

\bibitem{LevitinPRL09}
L.B. Levitin and T. Toffoli, {\em Phys. Rev. Lett.} {\bf 103}, 160502 (2009).

\bibitem{Nota_Caves}
The state
$|\alpha\rangle \otimes |\zeta \rangle$ 
($\otimes$ is the mode tensor product here)
reaches a SSN sensitivity
$\Delta \theta = \epsilon /\sqrt{|\alpha|^2}$, with $\epsilon < 1$ \cite{Caves_1981}.

\bibitem{nota_ex}
For the considered states written as $\ket{\psi}\equiv \sum_N \sqrt{Q_N}\ket{\psi^{(N)}}$, one can show that 
$\langle \hat J_{x,y}^{(N)}\rangle_{\ket{\psi^{(N)}}}=0$ holds for all $N$.
Eqs~(\ref{chiSV},\ref{xiSV})
are invariant since $F_Q(|\psi\rangle)=4(\Delta\hat J_y)^2_{\ket{\psi}}
=F_Q(\hat\rho_{\rm red})=\sum_N Q_N (\Delta\hat J_y^{(N)})^2_{\ket{\psi^{(N)}}}$ \cite{nota_FQ} 
holds and 
for block-diagonal operators $\mean{\hat A}_{\ket{\psi}}=\mean{\hat A}_{\hat\rho_{\rm red}}$.

\bibitem{convexity}
Convexity of $F$ has been proven in 
M. L. Cohen, {\em IEEE Transactions on Information Theory} {\bf 14}, 591 (1968).
A slightly more compact proof
is reported here.
For $\hat\rho(\theta)=\sum_k p_k \hat\rho_k(\theta)$ we have $P(\varepsilon|\theta)=
\tr[\hat E(\varepsilon) \hat\rho(\theta)] = 
\sum_k p_k P_k(\varepsilon|\theta) $
where $P_k(\varepsilon|\theta) = \tr[\hat E(\varepsilon) \hat\rho_k(\theta)]$.
Using the Cauchy-Schwarz inequality we have that 
$ (\partial P(\varepsilon|\theta) / \partial \theta)^2/ P(\varepsilon|\theta) \leq \sum_k p_k (\partial P_k(\varepsilon|\theta)/\partial \theta)^2/P_k(\varepsilon|\theta)$.
Integrating over $\ud \varepsilon$ we obtain that 
$F_{\hat E(\varepsilon)}[\sum_k p_k \hat\rho_k(\theta)] \leq \sum_k p_k F_{\hat E(\varepsilon)}[\hat\rho_k(\theta)]$.
This inequality holds for all possible POVM and, in particular, for those saturating the QFI.

\bibitem{UysPRA07} H. Uys and P. Meystre, \emph{Phys. Rev. A} \textbf{76}, 013804 (2007).


\end{thebibliography}
\end{document}